# Text Mining in Education

Rafael Ferreira[1], Máverick André, Anderson Pinheiro, Evandro Costa, Cristobal Romero

[1]Department of Computing, Federal Rural University of Pernambuco, Brazil

[2]Institute of Computing, Federal University of Alagoas, Brazil

[3]Department of Computer Sciences, University of Cordoba, Spain

**Correspondence**
Rafael Ferreira, Department of Computing, Federal Rural University of Pernambuco, Brazil
Email: rafael.mello@ufrpe.br

**Abstract:** The fast adoption of online education environments has in- creased in recent years, generating a large volume of data in text format from forums, chats, social networks, assess- ments, essays, among others. It has produced new chal- lenges on how to provide sophisticated mechanisms for ef- fective text data analysis in order to find relevant educa- tional knowledge. Despite the large number of educational application of text mining have been published recently, it was not found a survey detailing these applications. Thus, this paper presents an overview of the Educational Text Min- ing field from 2006 to 2018. It reviews this emerging area from three different viewpoints: the text mining techniques more used in educational environments, the most used edu- cational resources and the main applications or educational goals. Finally, it outlines the main conclusions, challenges, and future trends.

**K E Y W O R D S**
Educational Text Mining, Natural Language Processing in Education, Writing Analytics, Text Analytics

## 1 | INTRODUCTION

Currently, online education has become a new alternative to traditional education. In a broad sense, online education adopts online learning platforms for creating new educational settings able to support teachers and students in the learning process (Ring et al., 2013). Nowadays, there is a wide range of online education platforms, such as Course Management Systems (CMS), Learning Management System (LMS) and Massive Open Online Courses (MOOCs). These Virtual Learning Systems (VLE) engage hundreds or thousands of students that interact and generate a huge volume of structured and unstructured data requiring sophisticated methods of management and analysis (Romero and Ventura, 2017). In this context, the unstructured text data came from different sources in different formats such as discussion forum, chat, wiki, blogs, open questions, and essays, generally not suitable to be adequately processed and leading to problems related to using this information to can aid students and instructors in the learning process(Väljataga et al., 2011). To deal with the mentioned problems, traditional Educational Data Mining (EMD) and Learning Analytics (LA) techniques have been employed successfully for improving students' learning and aid teachers to manage the learning process (Wulf et al., 2014) (Slater et al., 2016). Although good results have been obtained, it does not fully explore all the educational resources available. For example, it is common to have learning activities involving open questions and essays exercises that could be used to evaluate the student and estimate the effectiveness of pedagogical strategies (Baker and Inventado, 2014). To address this issue, Text mining (TM) techniques could be adopted as it is in other areas such as risk management, fraud detection, business intelligence, and social media analysis. TM is the process to extract high-quality information from unstructured text (Berry and Castellanos, 2004). Hence, the new generation of online platforms could benefit from different text mining techniques such as:

- **Natural Language Processing**: It is the area of research that provides services to manipulate natural language text or speech (Chowdhury, 2003). It is largely used as preprocessing for others text mining techniques.
- **Text Classification and Clustering**: It applies the traditional classification and clustering algorithms to text data (Aggarwal and Zhai, 2012). The main difference is the preprocessing methods used to extract features from texts before the classification.
- **Information Retrieval**: It finds documents relevant for a specific query in large document databases (Manning et al., 2008). Recent methods also include question-answer systems.
- **Text Summarization**: It generates a compressed version of one or several documents containing the most important information (Gambhir and Gupta, 2017). The summarization could be extractive, where the summary is composed of sentences extracted from the text, or abstractive, where the methods identify the main content from text then it creates a summary.

In the educational domain, the TM has focused on analyzing the contents of educational resources (Kovanović et al., 2015), especially the research on educational text mining (Litman, 2016; Shum et al., 2016). TM techniques have achieved significant results especially in essays and forum analysis, academic text production and open questions evaluation (Lárusson and White, 2012; Simsek et al., 2015; Dascalu et al., 2015b). TM techniques could explore several educational resources, the most used are:

- **Forums and Chats**: These resources provide synchronous and asynchronous mechanisms for communication between teachers and students (Trausan-Matu et al., 2012; Lin et al., 2009; Kovanović et al., 2016).
- **Social Networks and Blogs**: The students use these resources to share their preferences, opinions and the relationships with other students. Thus, social networks and blogs are important to extract different information from students to improve recommendation systems and to avoid dropout (Jo et al., 2016; Ortigosa et al., 2014; Kechaou et al., 2011).
- **Online assignments and Essays**: They are the main resources of textual evaluation proposed in educational environments (Hsu et al., 2011; Dikli, 2006).

Despite the large number of educational applications of text mining have been published recently, it was not found a survey concerning these applications. Thus, this paper presents an overview of the Educational Text Mining (ETM) field, including the main techniques, resources, applications, and future trends. The research was performed from 2006 to July 2018, retrieving 343 relevant articles. More specifically, this paper intends to answer the following research questions:

*RQ1: What are the main text mining methods and techniques adopted in the educational technology field?*
*RQ2: What are the main educational sources and resources used for doing text mining?*
*RQ3: What are the main applications and educational goals?*

To answer these questions, this paper presents firstly the methodology applied to retrieval and select the papers used in the review in section 2. Then, section 3 describes the text mining methods and how they could be used to improve educational environments. Next, Section 4 presents educational resources and data sources which use TM to extract useful information. After that, Section 5 the main educational goals and applications that benefit from TM methods. Finally, Section 6 presents the final remarks and future trends of text mining and education.

## 2 | METHODOLOGY

This section presents details about the used method for selecting papers about the application of text mining to educational environments.

### 2.1 | Data Collection and Selection Criteria

Initially, we have searched using the keywords: "Text Mining in Education", "Natural Language Processing in Education" and "Writing Analytics" in the following databases: IEEE Xplore, Springer Link, ScienceDirect, ACM, and Google Scholar. All the papers were included in the first version of the analysis, and then we applied the next three exclusion criteria:

1. Publications where the search keywords did not appear in the title, abstract, conclusions, and keywords;
2. Publications involving text mining without educational goals or applications;
3. Publications that are not in the defined period (from January 2006 to July 2018).

After the previous exclusion process, we obtained 343 papers. Table 1 shows the most important academic databases of the papers. We can see that the three most important sources are: IEEE Xplore, Springer, and ACL.

### 2.2 | Data Classification and Analysis

In order to can classify the retrieved papers, we obtained some additional information from each paper such as the number of citations, year of publication and type of publication (journal, conference, book, and workshop).

As a summary, Figure 1 presents the number of papers published over the years. We can see that Only seven works were found in the first year of this study, although, the average number of publications per year is more than 26. However, reaching 56 only in the first seven months of 2018 that it shows the importance and growth of this research area in this last year.

About the type of publication, most of the papers were published in conference proceedings (59%), followed by 25% of journal papers, 15% of workshop papers and 1% of others, which include books and Ph.D. thesis (Figure 2). We can highlight some conferences and journals in relation to number of papers published in this field: Conference on Intelligent Tutoring Systems, Conference on Learning Analytics and Knowledge, Conference on Advanced Learning Technologies, Conference on Artificial Intelligence in Education; Conference on Educational Data Mining, and Journal of Artificial Intelligence in Education, Computers & Education, Expert Systems with Applications and Computers in Human Behavior.

**TABLE 1** Search Results - Academic Databases

| Database | Number of Papers (%) |
|---|---|
| Association for Computing Machinery (ACM) | 26 (7.58%) |
| Association for Computational Linguistics (ACL) | 51 (14.86%) |
| Educational Data Mining Society (EDMS) | 19 (5.53%) |
| Science Direct | 38 (11.07%) |
| IEEEXplorer | 70 (20.40%) |
| IGI Global | 11 (3.20%) |
| Springer | 55 (16.03%) |
| Others | 73 (21.28%) |

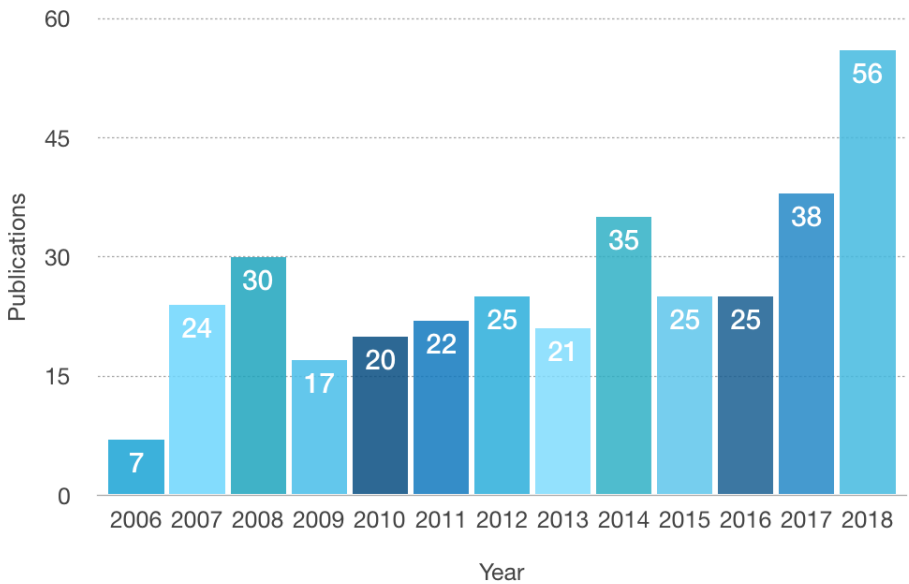

**FIGURE 1** Articles by Publication Year

About the number of citations, Table 2 shows the top-5 papers more cited according to Google Scholar. Three of them are reviews/surveys (De Wever et al., 2006) (Maurer et al., 2006) (Dikli, 2006) and the other two show experiments

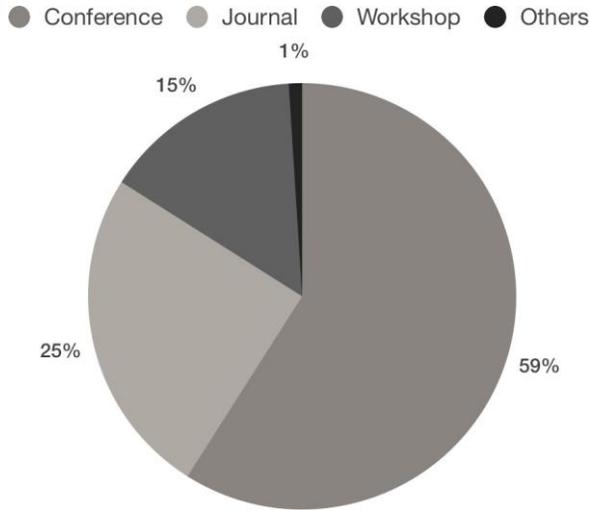

**FIGURE 2** Type of Publication

and applications (Baker et al., 2010) (Khribi et al., 2008).

**TABLE 2** *Top five most cited papers*

| Paper | Number of cites in GoogleSchoolar |
|---|:---:|
| (De Wever et al., 2006). | 1073 |
| (Baker et al., 2010) | 483 |
| (Maurer et al., 2006) | 374 |
| (Khribi et al., 2008). | 342 |
| (Dikli, 2006) | 337 |

Next, we categorized all the paper in three different ways: text mining methods, educational resources, and educational goals.

## 3 | TEXT MINING METHODS AND TECHNIQUES

This section introduces the text mining techniques and methods most used in education. The selected articles in this survey embraced different text mining techniques, as presented in Table 3. The main methods reported were text classification (32%), natural language processing (31%), theoretical (12%), information retrieval (5%), text clustering (5%), text summarization (4%) and others (12%). Besides, this research recovered theoretical papers, which present ideas about the application of text mining to education, and works related to different text mining applications (others) like information extraction, machine translate, text generation, among others.

**TABLE 3** *Publications Categories According to Text Mining Methods*

| Resource | No. of papers (%) |
|---|---|
| Text Classification | 109 (31.77%) |
| Natural Language Processing | 105 (30.62%) |
| Theoretical | 41 (11.95%) |
| Information Retrieval | 17 (4.96%) |
| Text Clustering | 17 (4.96%) |
| Text Summarization | 13 (3.79%) |
| Others | 41 (11.95%) |

## 3.1 | Text Classification and Clustering

Different machine learning algorithms have been used to extract information from texts. In general, these algorithms are divided into (Aggarwal and Zhai, 2012) classification or supervised learning, and clustering or unsupervised learning. Classification categorizes items based on their features in a predefined set of categories, and clustering categorizes items based on the similarity among them.

Machine learning could also be applied to different domains, including textual documents (Aggarwal and Zhai, 2012). The main difference between the processing of traditional data and text documents is the methods used to extract features from texts before the classification or clustering.

Educational environments apply text classification to different goals, for example: (i) the automatic classification of activities in three common discourse tasks: teacher lecturing, whole class discussion and student group work (Wang et al., 2014); and (ii) the categorization of forums discussion in topics (Tobarra et al., 2014; Azevedo et al., 2011) and genres (Lin et al., 2009). Despite these applications the ones that stand out in this field are: Sentiment Analysis (Newman and Joyner, 2018), Question Classification (Ruseti et al., 2018) and Automatic Scoring (Yoo and Kim, 2014). Especially, Sentiment Analysis is a new technique very used currently in education in order to determine the attitude of a speaker, writer, or another subject with respect to some topic or the overall contextual polarity or emotional reaction to a document, interaction, or event.

Regarding text clustering, the educational application found in the literature are related to group students and educational resources. Students clustering could be used to: improve text predictions (Trivedi et al., 2011), adapt the curriculum (Shi et al., 2015), measure engagement (Liu et al., 2013), identify learning patterns (Mansur and Yusof, 2013; Cobo et al., 2010), among others. On the other hand, educational resources are grouped in order to improve recommendation systems (Mansur and Yusof, 2013; Khribi et al., 2008). The SVM and K-means were the most used algorithms for text classification and clustering respectively. Besides, techniques based on neural networks are also largely used.

## 3.2 | Natural Language Processing

Natural Language Processing (NLP) is a field in computer science used to manipulate natural language text or speech (Chowdhury, 2003). It can process and analyze large amounts of natural language data by using algorithms for semantic and syntactic analysis, coreference resolution, named entity recognizer, among others.

Educational platforms largely adopt NLP techniques over the years; however, the growing of big-data, mobile technologies, social media, and MOOCs has resulted in the creation of many new research opportunities and challenges (Litman, 2016). These techniques have been applied to several educational resources, like essays and open question assessment, student feedback, forum/chat interactions, and automatic question generation.

The majority of applications of NLP to education are related to the automatic evaluation of essays and open questions. The literature proposes different methods to improve this kind of evaluation adopting text mining (Crossley et al., 2015). Regarding essays, NLP has been mainly used to analysis of text cohesion (Dascalu et al., 2015b; Balyan et al., 2017) and written argumentation (Persing and Ng, 2015; Elouazizi et al., 2017). For open questions, the leading research line is to semantically evaluate student's answers (Cutrone and Chang, 2010) and generate new questions (Flor and Riordan, 2018).

In addition, (Dzikovska et al., 2014) adopt different NLP techniques to provide feedback for intelligent tutoring based on student's interactions in basic electricity and electronics online courses. Following the idea of adopting NLP to enrich interactions, it was proposed papers to support collaborative work on chats (Trausan-Matu et al., 2012), predict group project performance (Yoo and Kim, 2014), and predict interactions based on past discussions (Kim and Shaw, 2014).

## 3.3 | Information Retrieval

Information Retrieval (IR) the science of searching for information in a document, searching for documents themselves, and also searching for metadata that describes data (Manning et al., 2008). Also, it is used to organize document facilitating navigation among them.

In the educational domain, different applications use IR to find relevant documents in libraries (Zhang and Gu, 2011; Chinkina et al., 2018). The main goal of these applications is to aid the student to find relevant books (Chen et al., 2008). It is also used to enhancing online discussion and collaboration in e-learning (Nuutinen and Sutinen, 2009; Distante et al., 2014).

As mentioned before, IR techniques improve the navigation among different texts. For educational purpose, it is used to visualize different topics in students essays and writing assignments (O'Rourke et al., 2011; Villalón and Calvo, 2011). Information retrieval could improve recommendation systems (Khribi et al., 2008; Mangina and Kilbride, 2008) and texts tagging systems (Vattam and Goel, 2011).

## 3.4 | Text Summarization

Automatic Text Summarization (ATS) creates a short version of one or several documents containing the most essential information (Gambhir and Gupta, 2017). It is mostly used to deal with the overload of information in text datasets, such as digital library, web news, scientific papers, among others.

ATS techniques are classified as Extractive and Abstractive. Extractive systems select a set of the most significant sentences from a document, exactly as they appear, to form the summary. Abstractive systems attempt to improve the coherence among the sentences in the summary by eliminating redundancies and clarifying their context.

Educational environments and resources apply ATS mainly to: (i) summarize contents (Mihalcea and Ceylan, 2007); (ii) extract keyphrase (Sándor and Vorndran, 2009); (iii) create visual models from student texts (Reategui et al., 2012); (iv) improve different applications (Jorge-Botana et al., 2015; Sung et al., 2016).

Besides the mentioned applications, ATS techniques are used to deal with traditional educational resources drawbacks, for example: (i) Automatic evaluation of forums posts (Azevedo et al., 2011); (ii) Assist students to write academic texts (Whitelock et al., 2015); Improve collaborative learning (Kang et al., 2008); (iii) Evaluate student feedback (Nitin et al., 2015).

# 4 | EDUCATIONAL SOURCES AND RESOURCES

This section details how all these previous text mining methods have been applied to education. We categorized all the selected papers in this survey according to the educational resource where the text mining was applied. Essays, forums, and online assignment achieved more than 50% of the retrieved works. It points the importance of these resources to educational environments. Besides, chats and social networks also reach a relevant number of published papers. Some papers present the motivation of the text mining application to education. In addition, the articles classified as Others include applications with blogs, wiki, course information, lecture notes, among others.

**TABLE 4** *Publication According to Its Educational Resources*

| Resource | No. of papers (%) |
|---|---|
| Online assignment | 66 (19.25%) |
| Essay | 61 (17.79%) |
| Forum | 55 (16.04%) |
| Chat | 29 (08.45%) |
| Document | 29 (08.45%) |
| Social Network | 10 (02.91%) |
| Others | 93 (27.11%) |

## 4.1 | Online assignments

An online assignment is a resource used to evaluate students progress as essays analysis. The two most popular type of assignments are: multiple-choice question (where there are some explicit answers to be selected) and open questions (where the student need to provide a free writing answer) (Wang et al., 2014). Text mining techniques support both the creation end evaluation of questions.

In order to automatically generate question and multiple-choice answers, Rajagopal *et al*. (Araki et al., 2016) proposed a method based on natural language processing and template-based algorithms. Mazidi and Tarau (Mazidi and Tarau, 2016) propose an evaluation of natural language understand and natural language generation methods to propose a new question.

Besides question creation, it is essential to analyze previous formulated questions. Text clustering and topic modeling techniques are applied to evaluate questions in a database qualitatively(Nagashree and Pujari, 2016). Besides, it is relevant to identify which question is relevant for each stage of learning (Sachan and Xing, 2016).

Concerning question evaluation, there are different proposals to aid the instructor in the correction of open questions(Cutrone and Chang, 2010; Rus et al., 2013; Rahimi et al., 2014). More recently, (Ruseti et al., 2018) proposed the adoption of deep learning methods to deal with the question quality problem.

The literature of this field also proposes semi-automatic methods to grading answer. Thus, the teacher benefits from o computational applications. However, he/she needs to evaluate the analytics (Hsu et al., 2011; Escudeiro et al., 2011).

## 4.2 | Essays

Differently, from previous resources, the primary goal of TM applied to Essay is to evaluate the students. Many exams, such as university admission and idiom proficiency, include an essay exercise. It evaluates the capacity of expression and critical insight from students (Knight et al., 2016). However, the task of analyzing an essay is complex, and some exams produce a large number of essays. Thus, it is important to create automatic techniques to aid the appraiser (Latif, 2008).

Text mining has been largely used to evaluate different aspects of essays automatically (Dikli, 2006; Rahimi et al., 2017). This field of research started by analyzing shallow features and error detection (Burstein, 2003). Errors in grammar, linguistic usage, and style are detected using statistical and natural language processing techniques (Crossley et al., 2015). Applications of text mining for essay analysis also include: Recommendation Systems (Acosta et al., 2014), User Modeling (Allen and McNamara, 2015), Plagiarism detection (Oberreuter and Velásquez, 2013), and Essay visualization model (Villalón and Calvo, 2011).

Another problem to assess essays is the discourse analysis (Warschauer and Ware, 2006). It benefits from syntactic structures and automatic discourse analysis methods to deal with it. Moreover, semantic features could be used to extract discourse information (Dascalu, 2014).

In writing assessments is important to measure the text cohesion. Recent literature proposed the adoption of different cohesion index for essays scoring (Crossley et al., 2013). Besides, (Vajjala, 2018) analyzed different linguistic features applied to non-native learner essays.

In general, the automatic scoring tools use a classifier to combine the outputs of each features mentioned above to present the final grade (Shermis et al., 2010). It is important to mention that different style of essays could require different analysis(Varner et al., 2013).

Besides the essay evaluation problem, text mining techniques could also be applied to help students to develop their writing skills. Thus, different approaches benefit from learning analytics and natural language processing for sending automatic feedback (McNamara et al., 2013; Lewkow et al., 2016; Woods et al., 2017).

Also, different works point that the interaction among students can improve the quality of text productions (Calvo et al., 2011). Southavilay *et al.* proposed different approaches based on text mining, maps and probabilistic topic models to support collaborative learning (Southavilay et al., 2013, 2010).

## 4.3 | Forum

A forum, or online discussion, is a communication tool where multiple users interact asynchronously. Thus, it is possible to make structured interventions over a specific time. In the educational context, the forum is the resource that allows greater interactivity between students and teachers (Caspi et al., 2003). Besides, it provides several possibilities for teachers to interact effectively with the class (De Wever et al., 2006). Despite the advantages created by adopting the forum as a communication resource in educational environments, it also led to an overload of information problem (Wulf et al., 2014). Therefore, it is important to propose automatic methods to extract relevant information from them.

Different works are exploring the text mining methods to provide user automatic feedback. Several papers propose the creation of conversational agents or chatter boots for educational forum (SHROFF and Deneen, 2011). These systems benefit from ontologies (Eisman et al., 2008), natural language processing (da Costa Pinho et al., 2013) and machine learning (Dzikovska et al., 2014) to interact with students trying to answer questions and help to understand the subject.

Besides, it is possible to use recommendation systems to provide user feedback. There are several works to suggest different resources for students. Some text-mining based approaches adopt Topic-Driven Content Search and similarities measures to recommend texts in an educational forum. The first approach proposes the application of information retrieval techniques to structure the searching and navigating in forum content for different topics of discussion (Distante et al., 2014). The second uses different similarities measures to find related questions in order to recommend an answer to the students (Catherine et al., 2012). The classification of posts, also called gender identification, is another topic widely researched in this field. Text classification could be used to assess students' online participation and contribution, investigation of an aspect of online learning instigated by a research project, evaluation, and monitoring of learning progress, among others (Lui et al., 2007). Another important application of text classification in this context is to avoid off-topic posts (Wu, 2017). Ravi and Kim (Ravi and Kim, 2007) propose the identification of student's interactions. Lin *et al.* (Lin et al., 2009) propose a system to classify post-gender using the words frequency as features and decision tree algorithm for classification. The works (Cao et al., 2011), and (Heiner and Zachary, 2009) use machine learning, graph representation and similarity measure to better modeling the student questions. Nunes *et al.* (Nunes et al., 2014) and WANG *et al.* (Wang et al., 2007) also intend to extract the context from forum. However, different from previous work, they provide the context all posts, not only the questions. Both papers combine semantic technologies and a statistical method to recommend relevant topics to be discussed in online discussion forums.

Also, to the post classification, the extraction of indicators from forums aid teachers to follow up the discussion. Mclaren *et al.* (McLaren et al., 2007) propose the use of Awareness Indicators to provide an interface that allows the teacher to oversee all of the e-discussions currently taking place in the classroom. Another relevant indicator is the cognitive presence, Joksimovic *et al.* (Joksimovic et al., 2014) automatically extracted a series of linguistic features to measure the student presence in forums. Cobo et al. (2010) propose an approach to generate a student forum activity model.

Besides the works presented, it is important to mention some works that highlight important educational features from forums, even though they do not propose any computational solutions. Murphy (Murphy, 2004) analyses 22 features related to the collaboration on forums. The author discusses the importance and frequency that they used to appear in the posts. Pena-Shaff and Nicholls (Pena-Shaff and Nicholls, 2004) propose a qualitative study to identify the main features related to knowledge constructions.

In addition, methods to incentive the student participation (Ghosh and Kleinberg, 2013) and sentiment analysis (Wen et al., 2014) are also relevant to educational forum success.

To conclude this section, it is important to notice that, although the interaction on an educational forum could assist students, the overlap posts may generate problems for understanding the subject. Thus, the teacher, and automatic systems need to identify the right moment to interact (Mazzolini and Maddison, 2007).

## 4.4 | Chats

Chat is a resource to provide synchronous communication. In educational environments, chats can be used to discuss a subject, perform group activity, ask questions, among others (Ferguson and Shum, 2011). The applications of text mining to chats are (Leiyue and Jianying, 2012): summarizing conversation topic; understanding the user behavior, investigation of chat user attributes, understanding social and semantic interactions, monitoring chat room conversations, extracting relevant information, authorship attribution and so on.

There are several works to identify and evaluate collaboration on educational chats. Trausan-Matu *et al*. (Trausan-Matu et al., 2012) propose a method using graph among utterances and words to identify threads and user contribution based on Bakhtin's dialogism (Koschmann, 1999). The collaboration could be automatically assessed using discourse cohesion analysis (Dascalu et al., 2015b). Another work that explores the collaboration is (da Costa Pinho et al., 2013). It proposes a pedagogical agent capable of mediating synchronous online discussions. Scheuer and McLaren (Scheuer and McLaren, 2008) adopt text mining for helping teachers to find important information in chat data. (Reimann et al., 2009) also provide information to help teacher by creating a model presenting the sequence of events discussed in chats. Trausan-Matu et al. (2012) propose the adoption of natural language processing techniques to find the most essential moments from student's discussion.

Besides, the papers (Rojas et al., 2012) and (Coutinho et al., 2016) introduce the sentiment analysis applied to a synchronous discussion. Finally, there are many papers working with the analysis of second language acquisition in a tutorial dialogue chat (Sinclair et al., 2018; Xu et al., 2018; Nayak and Rao, 2018)

## 4.5 | Documents, Social Networks, Blogs, and E-mails

Other resources that provides information about the interaction in educational environments are social networks, blogs, and emails. The text mining applications using these resources are usually related to text classification, mainly focusing on sentiment analysis. In this context, sentiment analysis intends to solve different problems, such as:

- Extract opinion about the educational environment (Kechaou et al., 2011). It is used to assist teachers to improve the educational environment increasing the student engagement.
- Create an adaptive educational environment (Ortigosa et al., 2014). This kind of environment tries to create a particular study guide for students, which in general improves their performance (Truong, 2016).
- Aid in teaching evaluation and feedback based on the user interaction on social network (Leong et al., 2012).

Different features could be applied to extract sentiment in educational platforms from traditional statistical features like TFIDF, Information Gain, Mutual Information and CHI statistics to natural language processing ones (Kechaou et al., 2011; Truong, 2016). It is important to notice that some social networks have a limited number of characters available for each post. In this case, it is important to perform a pre-processing step before extract features (Leong et al., 2012).

In addition to sentiment analysis, text classification is used to extract students' behavior (Tobarra et al., 2014). It aids teachers to improve feedback and prevent student dropout. Mansur and Yusof (Mansur and Yusof, 2013) proposed the adoption of a hybrid method based on a combination of logs from educational environments and social network analysis to extract user behavior. In the same direction, Tobarra et al. (2014) combines social network and forums interaction to generate students' models.

Another example of text mining application is regarding emails analysis (Aghaee, 2015).

Finally, it is also possible to automatically predict the grade of each student based on his/her interaction with the course blogosphere (Gunnarsson and Alterman, 2012).

# 5 | EDUCATIONAL GOALS AND APPLICATIONS

Text mining techniques have been applied to educational data with different goals (see Table 5) such as the extraction of useful information from educational forums and chats or the automatic evaluation of students' responses for essays and online assignments.

**TABLE 5** *Publication According to Its Educational Goal*

| Goal | No. of papers (%) |
| --- | --- |
| Evaluation | 95 (27.69%) |
| Student Support/Motivation | 48 (13.99%) |
| Analytics | 45 (13.11%) |
| Question/content Generation | 22 (06.41%) |
| User feedback | 18 (05.24%) |
| Recommendation Systems | 9 (02.62%) |
| Others | 106 (30.90%) |

We can see in Table 5 that the most popular application of text mining in education is for evaluation purposes, followed by student support and analytics.

## 5.1 | Evaluation

The TM methods have been largely used to evaluate student's performance in different contexts, especially to evaluate essays and online assignments (Kadupitiya et al., 2017; Dikli, 2006). Several works proposed the adoption of shallow features, as word counts, to assess essays (Dikli, 2006; Rudner et al., 2006). However, it is important to go beyond this analysis (Ericsson and Haswell, 2006; Crossley et al., 2015). Thus, recent papers focused on adoption of semantic methods (Simsek et al., 2015; Hughes et al., 2012), writing style (Oberreuter and Velásquez, 2013; Snow et al., 2015) and argumentation analysis Elouazizi et al. (2017).

Following a similar direction, the evaluation of online assignments adopts lexical and semantic approaches (Ramachandran and Gehringer, 2011; Prevost et al., 2012; Cutrone and Chang, 2010). Nevertheless, in this case, the works tend to be more focus on solving specific problems as plagiarism (Adeva et al., 2006), analyze short answer (Saha et al., 2018), and classify the questions (Godea et al., 2018).

Besides, it also could be applied in formative evaluation to assist educators to establish a pedagogical basis for decisions in order to maintain the environment (Lehman et al., 2012; Gibson et al., 2017) and evaluate interactions on educational online discussions (Yoo and Kim, 2014; Rubio and Villalon, 2016).

## 5.2 | Student Support/Motivation

Engaging the students in online platforms is essential. Especially in distance learning courses, the collaboration among students is essential for pedagogical success. Thus, TM is applied to provide support students (Murphy, 2004; Trausan-Matu et al., 2012; Liu et al., 2017a). Once again, the writing activity has the majority of publications on this goal. Several works focus on giving aid to students on the production of different text, such as traditional essays, academic documents, and e-books (Latif, 2008; O'Rourke and Calvo, 2009; Chen et al., 2007).

In addition, the analysis of interactional resources like forums, chats, and blogs are explored here. The application opinion analysis methods try to extract the sentiment of the students (Kechaou et al., 2011; Ortigosa et al., 2014; Sinclair et al., 2018), and the adoption of methods to encourage collaboration (Li et al., 2008; Trausan-Matu et al., 2012) are used to keep the students motivated preventing dropout. Moreover, there are a few works that aid the instructor to monitor students within this tools (Scheuer and McLaren, 2008; Perikos et al., 2017).

## 5.3 | Analytics

The text-based educational resources presented generate a large amount of content which makes it difficult to follow up them closely. Thus, it is vital to present analytics for professors continually improving the environment and providing feedback to students (Dzikovska et al., 2014; Lyons et al., 2018).

In this context, the TM methods have been adopted to extract information from activities such as writing production (Clemens et al., 2013) and forum/chat/e-mail (Ferguson and Shum, 2011; Aghaee, 2015). However, the majority of the applications extract information that could potentially help the instructor to analyze the students' performance and behavior.

In this case, the performance is extracted by using shallow features (Azevedo et al., 2011), a topic modeling approach (Nunes et al., 2014), and sentiment analysis (Tucker et al., 2014). On the other hand, the student different students' behavior and characteristics could be analyzed, for instance, Psychological characteristics, genres, and mental models (Joksimovic et al., 2014; Lintean et al., 2012; Lin et al., 2009). These results are applied for online scaffolding discussion, create groups, and prevent dropout.

## 5.4 | Question/content Generation and Application

TM has been applied to automatic generated questions and content to aid teachers in educational environments Araki et al. (2016); Mazidi and Tarau (2016). The most traditional approach is to generate the question from a reference document like textbooks and educational resources (Mazidi and Tarau, 2016; Araki et al., 2016), but it is also possible to find useful knowledge from web (Ochi and Nakanishi, 2009).

Another task classified in this goal was the scaffolding and evaluation of questions according to the course context. Sachan and Xing (2016) proposed a case study to analyze if the students are more likely to continue the course when the instructor send easier questions at first. In the same direction, (Becker et al., 2012) and (Mazidi and Nielsen, 2014) propose an algorithm to automatically ranking questions in different context and evaluation of automatically generating questions, respectively.

## 5.5 | User feedback

TM has also been used with the goal of providing automatic feedback for students based on both their interactions and activities performed (Lewkow et al., 2016; Woods et al., 2017). In general, the literature deals this goal using two different approaches.

The first one focuses on sending feedback directly to the students, as a dashboard or recommendation for example, based on an automatic method to analyze their behavior or performance (Lewkow et al., 2016). In this context, the range of applications goes from help students in a question-answering application (Alinaghi and Bahreininejad, 2011;

Zhu, 2015) to support students in collaborative environments (Dascalu et al., 2015a) and send feedback in real-time (Altrabsheh et al., 2017).

The second approach uses TM to extract information to aid instructors in the elaboration of feedback from different resources (Goldin et al., 2017). Gibson et al. (2017), Akçapınar (2015) and Hwang et al. (2007) presents different methods to help instructors to provide feedback based on writing activities. Also, several papers extract information to support formative feedback based on data from forums interactions and essays (Yang et al., 2011; Woods et al., 2017).

## 5.6 | Recommendation Systems

Due to the large amount of resource available in digital environments and on the Internet, it is difficult to find content to solve a doubt (Hsu et al., 2010; Mangina and Kilbride, 2008). So the adoption of recommendation systems aids the students in finding relevant information.

Their recommendations can be based on consolidated educational material as books (Nagata et al., 2009; Garrido et al., 2014), and on open environments like the Web (Sommer et al., 2014). Acosta et al. (2014) proposed a system that extracts relevant terms and keywords from the students' writings in order to search the web for related contents. In this case, the recommendation system filters the content more relevant for different students. Following the same idea Khribi et al. (2007) proposed the adoption of information retrieval methods to recommend content to students. The main difference is that it explores the learners' access history instead of their writings.

Another type of work within this goal is to organize repositories of content that can be potentially used in educational environments. In this context, Figueira (2008) uses text mining techniques to classify learning objects based on semantic categories in order to improve further recommendations.

## 5.7 | Others

Different educational application could benefit from TM, such as: automatic summarizing texts, visualization tools and curriculum adaptation. It also includes theoretical/motivational papers Reategui et al. (2012); Villalón and Calvo (2011); Shi et al. (2015).

## 6 | CONCLUSIONS AND FUTURE RESEARCH

Educational text mining has become increasingly adopted. Consequently, the scientific community is more interested in this field. This paper proposed a literature review of the application of text mining techniques in online education. A detailed search was carried out for papers between 2006 and July 2018. Its main contribution is to provide details about the main text mining techniques, educational resources, and goals proposed on the 353 relevant paper found in this research. In fact, after carried out our systematic review we can answer our three initial research questions:

*RQ1: What are the main text mining methods and techniques adopted in the educational technology field?*

As the previous sections described, the main text mining methods used in the educational literature are text classification and natural language processing, this result follows the same trend as the text mining techniques used in general applications (Hirschberg and Manning, 2015). However, the majority of educational text mining literature is more focused on the output than the process, which leads to systems with good accuracy, but with a lack of interpretation. The adoption of techniques as text extraction or text summarization, and increase the number of works regarding the feature engineering could generate in-deeper educational insights.

*RQ2: What are the main educational sources and resources used for doing text mining?*

The automatic evaluation of online assignments and essays together with the feedback and the seek for increasing of the interactive in forums led these resources to be extremely explored by the literature. Besides, the social networks, with the creation of communities applications, and the documents, especially for the recommendation of documents within digital libraries, also reached a good number of applications.

*RQ3: What are the main applications and educational goals?*

The main educational goals of TM applications deal with student evaluation in different resources. The automatic method to provide support to students was also an essential goal of TM. It could be reached by stimulating students' self-reflection or by presenting analytics to instructors that use it to assist them. In addition, the student feedback, for example, using recommendation systems and by aiding in collaborative writing and suggest correct answers, is a significant aspect explored by the works presented.

Finally, although, there is much work done, this research area still has some gaps to be improved, and there are also some hot and new topics to develop. For us after doing this review, the most interesting future researches lines are the next ones:

- **Online Discussion Collaboration and Participation**: Several papers try to address specific problems of online discussion (as seen in sections 4.3 and 4.4). However, a few of them deal with the participation and collaboration problems. The student participation in an educational environment, mainly using tools such as forum, can be decisive for the completion of the course, especially in distance education (Brinton et al., 2014). On the other hand, it is important to ensure not only the students' posts, but also the collaboration among them (Liu et al., 2017b). In virtual learning environments, collaboration among students is as important as the teacher's responses.
- **Writing Analytics**: In recent years, the Learning Analytics field has grown exponentially with applications to deal with different problems. It provides a set of measurement, analysis, and reporting of data about students in order to aid teachers and improve e-learning (Dawson et al., 2014). However, there is not much analytics available regarding text resources. Writing Analytics could improve teacher's feedback, the student monitoring, and evaluation (Shum et al., 2017; Gibson et al., 2017).
- **Natural Language Generation**: Although it has been used in some applications presented in this paper. Automatic generation of educational content is still an open challenge (Kempen, 2012). It could be applied to: generate online questions, interact with students in forums and chats, propose essays themes based on recent news, among others.
- **Text Mining for Different Languages**: One of the main problems in text mining field is to deal with different languages (Gupta et al., 2009). With the large adoption of online education, the environments need to be adaptable to different countries. Thus, it is essential to propose an application to deal with different languages. The educational data mining application does not have this problem because it uses data to extract information. It is one problem to widely adopted education text mining.
- **Sentiment Analysis**: Although nowadays there are a good number of works that use this technique (Newman and Joyner, 2018; Wen et al., 2014; Rojas et al., 2012; Coutinho et al., 2016; Ortigosa et al., 2014; Kechaou et al., 2011;

Sinclair et al., 2018; Tucker et al., 2014), this is a hot topic, and there is much work to do with it in the education field. On the one hand, it has to be applied to all educational resources and not only forum, chats, social networks, blogs, and e-mails, but also to essays and online assignments. Moreover, on the other hand, better and new algorithms to detect new different type of students' emotions has to be developed.

# 7 | ACKNOWLEDGEMENT

This research is supported by projects of the Spanish Ministry of Science and Technology TIN2017-83445-P.

Author One et al. | 25Whitelock, D., Twiner, A., Richardson, J. T., Field, D. and Pulman, S. (2015) Openessayist: a supply and demand learning analytics tool for drafting academic essays. In *Proceedings of the Fifth International Conference on Learning Analytics And Knowledge*, 208–212. ACM.

Woods, B., Adamson, D., Miel, S. and Mayfield, E. (2017) Formative essay feedback using predictive scoring models. In *Proceedings of the 23rd ACM SIGKDD International Conference on Knowledge Discovery and Data Mining*, 2071-2080. ACM.

Wu, S.-Y. (2017) Design of strategy for reducing off-topic in online discussion activities. In *Advanced Learning Technologies (ICALT), 2017 IEEE 17th International Conference on*, 184–185. IEEE.

Wulf, J., Blohm, I., Leimeister, J. M. and Brenner, W. (2014) Massive open online courses. *Business & Information Systems Engineering (BISE)*, **6**, 111-114.

Xu, S., Chen, J. and Qin, L. (2018) Cluf: a neural model for second language acquisition modeling. In *Proceedings of the Thirteenth Workshop on Innovative Use of NLP for Building Educational Applications*, 374–380.

Yang, Y., Heinrich, E. and Kemp, E. (2011) Assessment of online discussion: Collecting discussion content and generating analysis data for assessment and feedback. In *E-Learn: World Conference on E-Learning in Corporate, Government, Healthcare, and Higher Education*, 992-992. Association for the Advancement of Computing in Education (AACE).

Yoo, J. and Kim, J. (2014) Can online discussion participation predict group project performance? investigating the roles of linguistic features and participation patterns. *International Journal of Artificial Intelligence in Education*, **24**, 8-32.

Zhang, Y. and Gu, H. (2011) Text mining with application to academic libraries. *Computer science for environmental engineering and ecoinformatics*, 200-205.

Zhu, L. (2015) Engage students with a cloud-based student response system. In *E-Learn: World Conference on E-Learning in Corporate, Government, Healthcare, and Higher Education*, 51-51. Association for the Advancement of Computing in Education (AACE).